\documentclass{article}
\usepackage[T1]{fontenc}
\usepackage{spconf,amsmath,graphicx,hyperref}
\usepackage{amsfonts}
\usepackage{algorithmic}
\usepackage{algorithm}
\usepackage{array}
\usepackage[caption=false,font=normalsize,labelfont=sf,textfont=sf]{subfig}
\usepackage{textcomp}
\usepackage{stfloats}
\usepackage{url}
\usepackage{verbatim}
\usepackage{booktabs}
\usepackage{graphicx}
\usepackage{orcidlink}
\usepackage[nolist]{acronym}
\usepackage{cite}
\usepackage{multirow}
\usepackage{adjustbox}
\usepackage{pifont}
\usepackage[table]{xcolor}

\begin{acronym}
\acro{MAE}{mean absolute error}
\acro{CNN}{convolutional neural network}
\acro{CRNN}{convolutional recurrent neural network}
\acro{DoA}{direction of arrival}
\acro{DNN}{deep neural network}
\acro{DRR}{direct-to-reverberant ratio}
\acro{ELU}{exponential linear unit}
\acro{GMM}{Gaussian mixture model}
\acro{GRU}{gated recurrent unit}
\acro{IID}{interaural intensity difference}
\acro{ITD}{interaural time difference}
\acro{LSTM}{long short-term memory}
\acro{LPC}{linear predictive coding}
\acro{MSE}{mean squared error}
\acro{MLP}{multilayer perceptron}
\acro{RIR}{room impulse response}
\acro{RNN}{recurrent neural network}
\acro{SELD}{sound event localization and detection}
\acro{SNR}{signal-to-noise ratio}
\acro{SVM}{support vector machines}
\acro{STFT}{short-time fourier transform}
\end{acronym}

\newcommand{\meanci}[2]{#1\,{\scriptsize$\pm$#2}}
\definecolor{red_cool}{rgb}{0.5, 0.0, 0.0}
\title{Few-Shot Calibration for Sim-to-Real Single-Channel Speaker Distance Estimation}
\name{Michael Neri~\orcidlink{0000-0002-6212-9139}, Archontis Politis~\orcidlink{0000-0002-0595-2356}, Tuomas Virtanen~\orcidlink{0000-0002-4604-9729}}
\address{Faculty of Information Technology and Communication Sciences, Tampere University, Finland \\
\textit{\{michael.neri, archontis.politis, tuomas.virtanen\}@tuni.fi}}

\definecolor{Gray}{gray}{0.85}
\definecolor{red_cool}{rgb}{0.5, 0.0, 0.0}

\begin{document}
\ninept
\maketitle
\begin{abstract}
Speaker distance estimators are trained almost exclusively on simulated room acoustics, because real recordings annotated with the true talker-to-microphone distance are scarce. We show that models trained this way transfer poorly. On three real corpora we evaluate, simply predicting the average distance of the corpus is more accurate than any learned model. Then, we ask how few labelled real utterances are needed to make a frozen, synthetic-trained estimator useful, and study post-hoc calibration maps that rescale its output without gradients or retraining. An analysis of the achievable error shows that what the calibration is not limited by the absolute accuracy of the estimator, but how well it orders utterances by distance, since a constant bias or a wrong output scale is removed exactly by the calibration itself. Balancing this against the cost of estimating each coefficient from few samples yields a criterion that accounts for which map wins on which corpus and at which annotation budget, together with a shrinkage variant that requires no hard decision. Our findings suggest selecting synthetic checkpoints by linear correlation with true distances rather than by absolute error. Code, datasets, and analysis are available at \textit{\href{https://github.com/michaelneri/audio-distance-estimation}{https://github.com/michaelneri/audio-distance-estimation}}.
\end{abstract}

\section{Introduction}
\label{sec:intro}

Estimating the distance between a talker and a microphone from a single microphone is of practical interests for several applications, ranging from hearing aid devices~\cite{Hamacher_2005_EURASIP}, hands-free communication~\cite{Oh_1992_ICASSP}, and speech recognition~\cite{Omologo_1198_SpeechComm}. The cue is intrinsically acoustic: the direct-to-reverberant ratio, the early-reflection pattern and
the spectral envelope all vary systematically with distance~\cite{Neri_IWAENC_2026, zahorik2005auditory, mendoncca2016modeling}. When these pieces of information are degraded, both human and machines distance perception deteriorate considerably~\cite{cherry_1953_JournalAcousAme, Georganti_TASLP_2011}.

Early approaches to the more generic source distance estimation focused on binaural recordings and hand-crafted features derived from the \ac{DRR}~\cite{Lu_2010_TASLP} or from inter-channel correlations. These features were used to fit \acp{GMM} and \ac{SVM} classifiers to distinguish distance between discrete bins~\cite{Vesa_2007_WASPAA, Vesa_2009_TASLP, Georganti_2013_TASLP}. More recent studies revisited the problem with \acp{DNN}, yet most remain limited to binary far/near classification~\cite{Patterson_2022_Interspeech, krause2021joint} or to coarse distance bins over a narrow range~\cite{Georganti_TASLP_2011, Yiwere_2019_Sensor, Sobhdel_ICMLA_2024}. Yet, these methods typically required careful hyperparameter tuning and generalised poorly across different acoustic environments.

In fact, the practical obstacle is the availability of audio data with labeled distances. The research community has therefore converged on training with synthetic data. The work in~\cite{Neri_WASPAA_2023} was the first to frame speaker distance as a continuous regression problem, demonstrating that a \ac{DNN} operating on \ac{STFT} phase features can achieve centimeter-level accuracy in simulated environments. It is worth noting that the centimeter-level accuracy reported in~\cite{Neri_TASLP_2024, Neri_WASPAA_2023} was obtained on simulated data with time- and level-calibration. 
Similarly, in~\cite{GenDA2025_RoomAcoustics} the authors proposed a data augmentation approach for generating realistic room-acoustic conditions to support speaker distance estimation using the \ac{CRNN} model in~\cite{Neri_TASLP_2024}. Most directly relevant to this work, in~\cite{Neri_IWAENC_2026} it was shown that learning-based single-channel distance estimators rely predominantly on early reflections rather than other \ac{RIR} components, but it did not study the sim-to-real transfer for speaker distance estimation without the knowledge of the time calibration. The only result available in this scope is in~\cite{Neri_TASLP_2024}, which provided near-chance results on zero-shot experiments on three real corpora with a simulation-based estimator.

To mitigate this sim-to-real gap, the contributions of this work are:
(i) we carried out zero-shot speaker distance estimation analysis, analyzing how much mismatch is present when training on synthetic audios and testing on real data. (ii) We devised a \textit{few-shot calibration} that adapts a frozen simulation-based estimator to a new corpus with $N$ labelled utterances by fitting a two-parameter affine map, requiring no gradients and no retraining. (iii) Based on~\cite{Neri_IWAENC_2026}, we modified the head of the \ac{CRNN} to predict acoustic-based features of the enclosure. While this leaves simulation-based \ac{MAE} largely unchanged, it substantially improves correlation with true distance labels under domain shift, which is what governs how well the estimator can be calibrated. 

\section{Few-shot calibration}
\label{sec:method}

\subsection{Problem statement}

Let $f_\theta$ be a distance estimator trained exclusively on synthetic data $\mathcal{D}_s$ with learned parameters $\theta$ and kept frozen. Deployed on a target corpus $\mathcal{D}_t$ it predicts an uncalibrated distance $u = f_\theta(x)$, nominally in meters from the recording $x \in \mathbb{R}^{T_s}$ with $T_s$ samples. We assume access to a small calibration set $\mathcal{C}_N=\{(u_i,d_i)\}_{i=1}^{N}$ drawn from the training split of that corpus, with $N$ of the order of a few tens of samples, and find a map $g:\mathbb{R}^+\!\to\mathbb{R}^+$ producing a calibrated estimate $\hat{d}=g(u)$, requiring neither gradients, nor access to $\theta$, nor retraining. 

\subsection{Foundations of distance calibration}
\label{ssec:maps}

Two systematic effects displace $u$ from $d$. First, a regressor trained under a squared-error objective shrinks towards the prior mean of $\mathcal{D}_s$, whereas $\mathcal{D}_t$ concentrates its distances elsewhere, causing an \emph{offset}. Second, since the amplitude cue contributes negligibly \cite{Neri_IWAENC_2026}, the estimator maps early-reflection structure to distance under the room-geometry and absorption statistics seen in training; a target enclosure whose reverberant statistics differ therefore induces a multiplicative \emph{scale} error. Both can be corrected by the affine map
\begin{equation}
\hat{d} = a\,u + b ,
\label{eq:affine}
\end{equation}
with $(a,b)$ estimated by least squares on $\mathcal{C}_N$, without any domain adaptation or fine-tuning.

The constant predictor $\hat{d}=\bar{d}$ is the reference any calibration must beat: it consumes the same $N$ labels, estimates the average $\bar{d} = \mathbb{E}[\mathcal{C}_N]$, and ignores $f_\theta$ entirely. Constraining either coefficient of~\eqref{eq:affine} gives two one-parameter members of the same family, \emph{offset-only} ($a\!\equiv\!1$), which corrects the prior shift alone, and \emph{scale-only} ($b\!\equiv\!0$), which corrects the scale alone. Both trade modelling bias for estimation variance. 

\subsection{Estimating the affine mapping}
\label{ssec:rule}

Let $\mu_d,\sigma_d$ and $\mu_u,\sigma_u$ denote the target-corpus label and raw predicted distance statistics, and $\rho$ the Pearson correlation between $u$ and $d$. Theoretically with unlimited number of samples ($N \rightarrow +\infty$), the coefficients of~\eqref{eq:affine} that minimise the population risk $R(a,b)=\mathbb{E}[(d-\hat{d})^2] = \mathbb{E}[(d-au-b)^2]$ are obtained by solving $\partial R(a,b)/\partial a = 0$ and $\partial R(a,b)/\partial b = 0$, yielding the optimal coefficients
\begin{equation}
b^\star = \mu_d - a^\star\mu_u ,
\qquad
a^\star = \rho\,\frac{\sigma_d}{\sigma_u} ,
\label{eq:opt}
\end{equation}
Substituting back to $R(a,b)$ yields after some calculations
\begin{equation}
\mathbb{E}\big[(d-\hat{d})^2\big] = \sigma_d^2\,(1-\rho^2).
\label{eq:oracle}
\end{equation}
It is worth noting that $\rho$ is invariant to affine transformations of $u$, so a constant bias or a wrong output scale leaves~\eqref{eq:oracle} unchanged. What remains is how well the model correlates with distances within the target corpus. A model with large zero-shot \ac{MAE} and high $\rho$ calibrates better than an accurate but unordered one. With finite $N$, $(a,b)$ are themselves estimated from $\mathcal{C}_N$, and their sampling error adds to the test risk. An affine map ($p=2$ free parameters) has expected test risk

\begin{equation}
R_N^{\mathrm{aff}} \;\approx\; R(a^{\star},b^{\star})\Big(1+\frac{2}{N}\Big)
 = \sigma_d^2\,(1-\rho^2)\Big(1+\frac{2}{N}\Big),
\label{eq:finiteN}
\end{equation}

The constant predictor is the special case $a=0$, so the same reasoning applies with $p=1$,
\begin{equation}
R_N^{\mathrm{const}} \;\approx\; R(0,\mu_d)\Big(1+\frac{1}{N}\Big)
 = \sigma_d^2\Big(1+\frac{1}{N}\Big).
\label{eq:finiteN0}
\end{equation}

Having both the risk of the affine and the constant calibrations, we can estimate when it is worth fitting a slope, i.e., when $R_N^{\mathrm{aff}} < R_N^{\mathrm{const}}$, which yields the inequality
\begin{equation}
\rho^2 \;>\; \frac{1}{N+2}.
\label{eq:rule}
\end{equation}
Using this inequality which includes the number of calibrating samples and the correlation between ground-truth and predicted distances of the target corpus, we can set the threshold in function of the cardinality of the calibration set, i.e.\ $\rho>0.38$ at $N\!=\!5$, $0.29$ at $N\!=\!10$ and $0.21$ at $N\!=\!20$. When the inequality does not hold, the least-squares slope is dominated by its own sampling noise. 

Offset-only admits an analogous test. Shifting the estimator without rescaling it leaves the score spread intact, so its risk is $R(1,\mu_d-\mu_u)=\sigma_d^2+\sigma_u^2-2\rho\sigma_d\sigma_u$. As it estimates a single parameter, like the constant predictor, both carry the same variance penalty and the $(1+1/N)$ factors cancel; the comparison therefore involves no $N$ at all and reduces to
\begin{equation}
\sigma_u/\sigma_d \;<\; 2\rho ,
\label{eq:offrule}
\end{equation}
which holds only when the estimator \emph{under}-disperses relative to the target labels.

It is worth noting that the correlation between the estimator and the calibration set is itself estimated from few samples, and thus noisy at small $N$. We denote this estimate $\hat{\rho}$. Then, we propose an additional mapping that avoids a hard threshold on~\eqref{eq:rule} and instead shrinks the slope in proportion to the evidence for it~\cite{efron1973stein},

\begin{equation}
\hat{a}_\lambda = \frac{\hat{\rho}^2}{\hat{\rho}^2 + 1/N}\;\hat{a},
\qquad
\hat{b}_\lambda = \bar{d} - \hat{a}_\lambda\,\bar{u},
\label{eq:shrunk}
\end{equation}
which reduces to the constant predictor as $\hat{\rho}\!\to\!0$, restores \eqref{eq:affine} as $N\!\to\!\infty$, and halves the slope at $\hat{\rho}^2=1/N$, that is a smoothed form of~\eqref{eq:rule}.

\section{Materials}
\label{sec:results}

\subsection{Distance estimator}
We use as distance estimator the convolutional-recurrent regressor of~\cite{Neri_WASPAA_2023} as the baseline. It processes \ac{STFT} log-magnitude and sine/cosine phase features, a three-block \ac{CNN} with max/average pooling, a two-layer bidirectional \ac{GRU}, and a per-frame distance head pooled over time. Input is $10$\,s of mono audio at $16$\,kHz. We additionally evaluate a full-stack variant that adds, over the baseline, (i) auxiliary regression heads that predict $\log T_{60}$, $\log T_{\mathrm{mix}}$, and $\log V$ from the mean-pooled recurrent features, adding roughly $100$k learnable parameters. Here $T_{\mathrm{mix}}$ is the mixing time, the boundary between early and late reflections in the \ac{RIR}, obtained from the echo-density measure of~\cite{abel2006}. These parameters are chosen because they summarise the reverberant statistics that carry the distance cue; (ii) SpecAugment~\cite{Park_2019_Interspeech} ($2$ time masks, $2$ frequency masks) applied after feature extraction; and (iii) waveform augmentation (polarity inversion, $\pm6$\,dB gain, $\pm5$\,ms shift, each with probability $0.5$). Both variants are trained with Adam at $10^{-3}$ under two regimes: \emph{clean-trained}, and \emph{noise-trained} with WHAM!~\cite{Wichern_2019_Interspeech} noise mixed in at randomised \ac{SNR} spanning uniformly $0-50$ dB. Following~\cite{Neri_TASLP_2024}, both the clip-level prediction $\hat{y}$ and the per-frame prediction $\hat{\mathbf{y}}_t$ are supervised, which encourages the recurrent layers to produce sharper per-frame estimates, giving
\begin{equation}
\mathcal{L} = \tfrac{1}{2}\bigl[\ell(\hat{y},y)+\ell(\hat{\mathbf{y}}_t,\mathbf{y}_t)\bigr]
            + \lambda\!\!\sum_{k\in\mathcal{K}}\!\ell(\hat{u}_k,\log k),
\label{eq:loss}
\end{equation}
where $\ell$ is the \ac{MSE}, $\mathcal{K}=\{T_{60},T_{\mathrm{mix}},V\}$ collects the auxiliary targets with predictions $\hat{u}_k$ and weight $\lambda=0.3$, and $y\in\{\log d,d\}$ depending on whether the log-distance target is enabled. Auxiliary terms are dropped when the corresponding head is disabled in an ablation. All variants use five-fold cross-validation, giving five checkpoints each.

\subsection{Synthetic dataset}
\label{ssec:dataset}
We employ the same uncalibrated dataset (neither time- nor amplitude-calibration) as in~\cite{Neri_IWAENC_2026}. Specifically, anechoic speech recordings obtained from the EARS dataset~\cite{Richter_Interspeech_2024} are convolved with the simulated \acp{RIR} from \textit{pyroomacoustics}~\cite{Scheibler_ICASSP_2018}. The experiments include $2500$ audio files of $10$ s duration at $16$ kHz. The samples are randomly assigned to $5$ folds to assess the performance in a $5$-fold cross-validation fashion. By doing so, each cross-fold iteration assigns $1500$, $500$, and $500$ audios to training, validation, and testing sets, respectively. We follow the five-fold cross-validation protocol of~\cite{Neri_IWAENC_2026}: for each fold $i\!\in\!\{0,\dots,4\}$, fold $i$ is held out for testing, fold $(i{+}1)\bmod 5$ for validation, and the remaining three folds are used for training. The simulated rooms span a wide acoustic range: source-to-microphone distances from $1.00$ to $11.00$\,m (mean $5.8$\,m), reverberation times $T_{60}$ from $0.28$ to $2.16$\,s (mean $0.77$\,s), room volumes from $87$ to $883\,\text{m}^3$ (mean $441\,\text{m}^3$), and mixing times $T_{\text{mix}}$ from $84$ to $166$\,ms (mean $129$\,ms).

\subsection{Real datasets}
We evaluate sim-to-real transfer on three corpora spanning two levels of realism. \textbf{VoiceHome2}~\cite{Bertin_2019_SpeechCommunication} which encompasses smart-home commands recorded by twelve speakers in twelve rooms across four houses, under quiet and noisy conditions with uncontrolled domestic interferers (competing talkers, TV, appliances) and no \ac{SNR} annotation. Five source positions per room, standing and sitting, with a fixed 8-microphone MEMS array on a cubic baffle where we use the first channel only. In total, the dataset encompasses $752$ recordings of $10$ s, with distances spanning $1.0$-$4.5$ m (mean $2.27$ m). \textbf{STARSS23}~\cite{shimada2023starss23}. Multi-speaker interaction scenes recorded at Tampere University and Sony in eleven rooms with an Eigenmike array, from which a single omnidirectional channel is extracted. We use $2934$ single-speech excerpts that do not overlap with other annotated directional sources. The corpus is the most challenging of the three: speakers move and change orientation, and diffuse and directional ambient noise is present at significant levels. Distances span $1.5$–$2.9$ m (mean 2.19 m). \textbf{QMUL-TIMIT}~\cite{Neri_TASLP_2024}. Measured omnidirectional \acp{RIR} captured in three rooms at Queen Mary University of London: a 7.5×9×3.5 m classroom ($236$ m$^3$, $130$ \acp{RIR}), the Octagon, an eight-walled Victorian hall with a $21$ m domed ceiling ($\approx 9500$ m$^3$, $169$ \acp{RIR}), and the Great Hall ($169$ RIRs over a 12×12 m region). Each \ac{RIR} is convolved with $5$ anechoic TIMIT utterances, yielding $2340$ recordings, with \acp{RIR} split $70$/$10$/$20$ into training/validation/testing. We report clean and $0$ dB conditions with WHAM! noise, matching the synthetic protocol.

\section{Results}

\subsection{Results on synthetic data}

Table~\ref{tab:main_grid} reports the \ac{MAE} at the clean and $0$\,dB extremes. Training the baseline with noise already recovers most of the $0$\,dB degradation ($1.83 \to 1.47$\,m); on top of the noise-aware regime the multi-task and augmentation stack gives a smaller further gain ($1.47 \to 1.41$\,m) while also improving the clean condition ($1.30 \to 1.23$\,m). 
The per-component contributions are not monotonic: the log-distance target alone significantly \emph{increases} $0$\,dB \ac{MAE} under clean training ($p<0.05$), and neither the multi-task heads nor SpecAugment alone produce a significant change. A plausible explanation is that the log-distance loss reweights gradients toward small distances, which sharpens the target but offers no benefit without the implicit regularisation that noise diversity provides; under noise-aware training the same target is harmless. Only the full stack yields a significant improvement in both regimes.

\begin{table}[th]
\caption{Test average \ac{MAE} in meters with $95\%$ confidence half-widths (5 cross-validation folds, $t$-distribution) at the two extreme \ac{SNR} conditions. Best per column in \textbf{bold}. Markers give the paired $t$-test against the baseline at $0$\,dB: $^{*}p<0.05$, $^{**}p<0.01$. The \emph{Random} row predicts the training-set mean, serving as a sanity check.}
\label{tab:main_grid}
\centering
\footnotesize
\setlength{\tabcolsep}{4pt}
\adjustbox{max width=\columnwidth}{%
\begin{tabular}{@{}l cc cc@{}}
\toprule
& \multicolumn{2}{c}{\textbf{Clean-trained}} & \multicolumn{2}{c}{\textbf{Noise-trained}} \\
\cmidrule(lr){2-3}\cmidrule(lr){4-5}
\textbf{Model} & clean & $0$\,dB & clean & $0$\,dB \\
\midrule
Baseline~\cite{Neri_IWAENC_2026}   & \meanci{1.30}{0.08} & \meanci{1.83}{0.17} & \meanci{1.29}{0.06} & \meanci{1.47}{0.11} \\
\quad + log-distance target        & \meanci{1.38}{0.13} & \meanci{2.01}{0.14}$^{*}$ & \meanci{1.31}{0.09} & \meanci{1.46}{0.13} \\
\quad + multi-task heads           & \meanci{1.35}{0.06} & \meanci{1.92}{0.19} & \meanci{1.34}{0.12} & \meanci{1.51}{0.10} \\
\quad + SpecAugment                & \meanci{1.32}{0.13} & \meanci{1.78}{0.20} & \meanci{1.29}{0.12} & \meanci{1.49}{0.11} \\
\quad + speech aug.\ (full stack)  & \textbf{\meanci{1.23}{0.13}} & \textbf{\meanci{1.69}{0.13}}$^{**}$ & \textbf{\meanci{1.23}{0.09}} & \textbf{\meanci{1.41}{0.11}}$^{*}$ \\
\midrule
Random & \multicolumn{4}{c}{\meanci{2.47}{0.06}} \\
\bottomrule
\end{tabular}}
\end{table}

\subsection{Zero-shot real data analysis}

We revisit zero-shot analysis proposed in~\cite{Neri_TASLP_2024} using the same two real corpora (VoiceHome2~\cite{Bertin_2019_SpeechCommunication} and STARSS23~\cite{shimada2023starss23}) and the hybrid QMUL-TIMIT~\cite{Neri_TASLP_2024} to analyze the sim-to-real gap without any fine-tuning. For the comparison, we report two baselines that draw the label-prior floor. The first, \textit{Random (synth.)}, predicts the synthetic training-set mean for every sample, without requiring any real-data labels. The second, \textit{Random (real)}, predicts the training-split label mean of each target corpus. However, it is worth noting that the \textit{Random (real)} baseline is not strictly zero-shot as it requires label access to the target domain. 

\begin{table}[hbt!]
  \centering
  \caption{Zero-shot real-data test \ac{MAE} in meters (mean\,$[lo,\,hi]$ 95\,\% CI) and Pearson correlation $\rho$ per corpus, across five fold checkpoints ($^{*}p<0.05$, $^{***}p<0.001$). QMUL-TIMIT (real measured \acp{RIR}) is reported \emph{clean} and at $0$\,dB \ac{SNR}. \emph{Random} baselines are constant predictors ($\rho{=}0$ by definition). Best learned model per corpus in \textbf{bold}.}
  \label{tab:real}
  \small
  \setlength{\tabcolsep}{4pt}
  \adjustbox{max width=\columnwidth}{
  \begin{tabular}{@{} l cc cc @{}}
    \toprule
    \textbf{Model} & \textbf{VoiceHome2} & \textbf{STARSS23} & \multicolumn{2}{c}{\textbf{QMUL-TIMIT}} \\
    \cmidrule(lr){4-5}
    & & & \textbf{clean} & \textbf{0\,dB} \\
    \midrule
    \multirow{2}{*}{Baseline, clean-trained}
      & $2.38\;{\scriptstyle [1.66,\,3.11]}$
      & $3.07\;{\scriptstyle [2.28,\,3.86]}$
      & $\mathbf{5.97}\;{\scriptstyle [5.36,\,6.58]}$
      & $5.54\;{\scriptstyle [5.02,\,6.05]}$ \\
    & $(\rho{=}{+}0.08^{*})$ & $(\rho{=}{+}0.09^{***})$ & $(\rho{=}{+}0.18^{***})$ & $(\rho{=}{+}0.01)$ \\
    \addlinespace
    \multirow{2}{*}{Baseline, noise-trained}
      & $2.04\;{\scriptstyle [1.86,\,2.22]}$
      & $2.79\;{\scriptstyle [2.32,\,3.25]}$
      & $7.30\;{\scriptstyle [6.57,\,8.03]}$
      & $\mathbf{4.84}\;{\scriptstyle [4.60,\,5.07]}$ \\
    & $(\rho{=}{+}0.01)$ & $(\rho{=}{+}0.15^{***})$ & $(\rho{=}{+}0.17^{***})$ & $(\rho{=}{+}0.05^{*})$ \\
    \addlinespace
    \multirow{2}{*}{Full stack, clean-trained}
      & $1.91\;{\scriptstyle [1.35,\,2.47]}$
      & $1.57\;{\scriptstyle [1.19,\,1.95]}$
      & $7.66\;{\scriptstyle [7.04,\,8.28]}$
      & $6.18\;{\scriptstyle [5.63,\,6.73]}$ \\
    & $(\rho{=}{+}0.08)$ & $(\rho{=}{+}0.14^{***})$ & $(\rho{=}{+}0.55^{***})$ & $(\rho{=}{+}0.28^{***})$ \\
    \addlinespace
    \multirow{2}{*}{Full stack, noise-trained}
      & $\mathbf{1.79}\;{\scriptstyle [1.63,\,1.96]}$
      & $\mathbf{1.51}\;{\scriptstyle [1.13,\,1.88]}$
      & $8.32\;{\scriptstyle [7.84,\,8.81]}$
      & $5.62\;{\scriptstyle [5.14,\,6.09]}$ \\
    & $(\rho{=}{+}0.05)$ & $(\rho{=}{+}0.18^{***})$ & $\mathbf{(\rho{=}{+}0.63^{***})}$ & $\mathbf{(\rho{=}{+}0.29^{***})}$ \\
    \midrule
    Random(synth)$^{\star}$     & $3.68$ & $3.81$ & $4.57$ & $4.57$ \\
    \rowcolor{gray!20}
    Random(real)$^{\star\star}$ & $0.80$ & $0.45$ & $3.01$ & $3.01$ \\
    \midrule
    \multicolumn{5}{l}{\parbox{\columnwidth}{\scriptsize
      $^{\star}$Predicts the synthetic training-set mean (6.0\,m).\\
      $^{\star\star}$Predicts the training-split label mean of each corpus ($2.27$ m, $1.76$ m, $8.95$ m for VoiceHome2, STARSS23, QMUL-TIMIT). Requires real-data label access; included as a label-prior lower bound.}}\\
    \bottomrule
  \end{tabular}
  }
\end{table}

\begin{figure}[t]
    \centering
    \includegraphics[width=0.83\columnwidth]{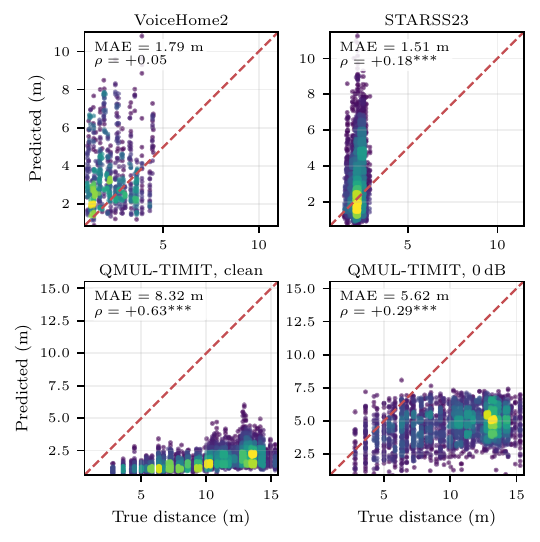}
    \caption{Zero-shot predictions of the full-stack noise-trained model on the four real test conditions. The dashed line is the identity; colour encodes point density. Correlations are Pearson, as in Table~\ref{tab:real} ($^{***}p<0.001$).}
    \label{fig:scatter_realdata}
\end{figure}

Results of the zero-shot analysis are shown in Table~\ref{tab:real}. On VoiceHome2 and STARSS23 every learned model outperforms the synthetic constant predictor ($3.68$\,m and $3.81$\,m), confirming that some transferable distance cue survives the domain shift. On QMUL-TIMIT the ordering reverses: \textit{Random(synth)} reaches $4.57$\,m while the learned models score $5.97$--$8.32$\,m, despite correlating distances far better. On VoiceHome2 the correlation never exceeds $+0.08$ and is significant in only one of four configurations, so the \ac{MAE} gain over the constant predictor reflects a distribution shift rather than genuine distance correlation. This is not a range effect: STARSS23 spans a strictly narrower interval ($\sigma_d{=}0.26$\,m against $0.95$\,m) yet reaches $\rho{=}{+}0.18$. On STARSS23 all models achieve significant correlations ($\rho = 0.09$--$0.18$, $p<0.001$), the full-stack noise-trained model attaining $\rho{=}{+}0.18$ alongside the lowest \ac{MAE} of $1.51$\,m, so even in this regime the full-stack architecture extracts a weak but reliable distance signal. The top row of Fig.~\ref{fig:scatter_realdata} shows what this looks like: predictions form near-vertical smears with little dependence on the true distance.

\begin{table}[t]
  \centering
  \caption{Calibration \ac{MAE} in meters for every map of Section~\ref{sec:method}, full-stack noise-trained model. \emph{Uncalibrated} is the zero-shot \ac{MAE} of Table~\ref{tab:real}. $N$ labelled samples are drawn from the target training split (mean over 500 draws); evaluation on the full test split. Best per column in \textbf{bold}. Markers denote a significant improvement over the constant predictor (paired $t$-test across folds): $^{*}p<0.05$, $^{**}p<0.01$, $^{***}p<0.001$.}
  \label{tab:variants}
  \small\setlength{\tabcolsep}{4pt}
  \adjustbox{max width=0.65\columnwidth}{
  \begin{tabular}{@{} l cccc @{}}
    \toprule
    \textbf{Calibration map} & $N{=}5$ & $N{=}10$ & $N{=}20$ & $N{=}50$ \\
    \midrule
    \multicolumn{5}{@{}l}{\textit{VoiceHome2}}  \\
    \quad Uncalibrated ($\hat{d}{=}u$) & \multicolumn{4}{c}{1.79} \\
    \quad Constant $\bar{d}$ & \textbf{0.85} & \textbf{0.82} & \textbf{0.81} & \textbf{0.80} \\
    \quad Offset-only ($a{=}1$) & 1.69 & 1.62 & 1.58 & 1.56 \\
    \quad Scale-only ($b{=}0$) & 1.17 & 1.12 & 1.09 & 1.07 \\
    \quad Affine~\eqref{eq:affine} & 1.00 & 0.87 & 0.83 & 0.81 \\
    \quad Shrunk affine~\eqref{eq:shrunk} & 0.92 & 0.85 & 0.82 & \textbf{0.80} \\
    \addlinespace
    \multicolumn{5}{@{}l}{\textit{STARSS23}} \\
    \quad Uncalibrated ($\hat{d}{=}u$) & \multicolumn{4}{c}{1.51} \\
    \quad Constant $\bar{d}$ & \textbf{0.47} & \textbf{0.45} & 0.45 & 0.45 \\
    \quad Offset-only ($a{=}1$) & 2.01 & 1.93 & 1.90 & 1.88 \\
    \quad Scale-only ($b{=}0$) & 1.10 & 1.11 & 1.12 & 1.12 \\
    \quad Affine~\eqref{eq:affine} & 0.51 & 0.46 & \textbf{0.44}$^{*}$ & \textbf{0.44}$^{**}$ \\
    \quad Shrunk affine~\eqref{eq:shrunk} & 0.49 & \textbf{0.45} & \textbf{0.44}$^{*}$ & \textbf{0.44}$^{**}$ \\
    \addlinespace
    \multicolumn{5}{@{}l}{\textit{QMUL-TIMIT, clean}} \\
    \quad Uncalibrated ($\hat{d}{=}u$) & \multicolumn{4}{c}{8.32} \\
    \quad Constant $\bar{d}$ & 3.20 & 3.08 & 3.05 & 3.02 \\
    \quad Offset-only ($a{=}1$) & 2.88$^{**}$ & 2.78$^{**}$ & 2.71$^{**}$ & 2.69$^{**}$ \\
    \quad Scale-only ($b{=}0$) & \textbf{2.79}$^{**}$ & 2.60$^{**}$ & 2.52$^{**}$ & 2.47$^{**}$ \\
    \quad Affine~\eqref{eq:affine} & 3.01$^{*}$ & \textbf{2.54}$^{***}$ & \textbf{2.39}$^{***}$ & \textbf{2.30}$^{***}$ \\
    \quad Shrunk affine~\eqref{eq:shrunk} & 2.91$^{**}$ & 2.59$^{***}$ & 2.46$^{***}$ & 2.33$^{***}$ \\
    \addlinespace
    \multicolumn{5}{@{}l}{\textit{QMUL-TIMIT, $0$\,dB}}  \\
    \quad Uncalibrated ($\hat{d}{=}u$) & \multicolumn{4}{c}{5.62} \\
    \quad Constant $\bar{d}$ & 3.19 & 3.09 & 3.05 & 3.02 \\
    \quad Offset-only ($a{=}1$) & \textbf{2.88}$^{***}$ & \textbf{2.76}$^{***}$ & \textbf{2.70}$^{***}$ & \textbf{2.67}$^{***}$ \\
    \quad Scale-only ($b{=}0$) & 3.17 & 2.99$^{**}$ & 2.88$^{**}$ & 2.82$^{***}$ \\
    \quad Affine~\eqref{eq:affine} & 3.78 & 3.07 & 2.82$^{**}$ & 2.70$^{**}$ \\
    \quad Shrunk affine~\eqref{eq:shrunk} & 3.37 & 2.99 & 2.82$^{**}$ & 2.72$^{**}$ \\
    \bottomrule
  \end{tabular}}
\end{table}

On QMUL-TIMIT the evaluation separates correlation from absolute calibration, as the bottom row of Fig.~\ref{fig:scatter_realdata} illustrates. Under clean conditions the full-stack model achieves $\rho{=}0.63$ ($p{<}0.001$), far above the baseline ($\rho{=}0.17$), despite a much higher absolute \ac{MAE} ($8.32$\,m vs.\ $7.30$\,m): the predictions form a tight, clearly ordered band lying well below the identity line, so the model is badly miscalibrated in scale relative to QMUL-TIMIT's distance distribution while preserving the correct ordering. At $0$\,dB the contrast persists, as the baselines retain essentially no correlation ($\rho \leq {+}0.05$) while the full-stack model reaches $\rho{=}{+}0.29$ ($p{<}0.001$), showing that the full-stack recipe is markedly more robust under domain shift.

\subsection{Few-shot calibration}
We draw $N$ samples from the training split of each corpus, fit every map of Section~\ref{sec:method} by least squares, and evaluate on the full test split; Table~\ref{tab:variants} reports the mean over $500$ draws for the full-stack noise-trained model. No single map dominates, and the winner is set by the variance budget of~\eqref{eq:rule} together with the spread ratio of~\eqref{eq:offrule}. On VoiceHome2 and STARSS23 the estimator over-disperses ($\sigma_u/\sigma_d = 1.8$ and $4.4$) while $\rho\!\approx\!0$, so both conditions fail: on VoiceHome2 \emph{every} map is significantly worse than the constant predictor at every $N$ ($p<0.01$), and on STARSS23 the affine map only overtakes it from $N{=}20$ and then by $0.01$\,m. Offset-only, which retains the inflated spread, is the worst variant on both ($1.69$ and $2.01$\,m at $N{=}5$). On QMUL-TIMIT the estimator under-disperses ($0.17$ clean, $0.28$ at $0$\,dB) and the picture reverses: offset-only significantly beats the constant predictor at every $N$ in both conditions, winning every column at $0$\,dB and beating the affine map even at $N{=}50$ ($2.67$ vs.\ $2.70$\,m). The affine map leads only where $\rho$ is large enough to pay for its second parameter, on QMUL-TIMIT clean from $N{=}10$ onwards ($2.54$\,m against $3.08$\,m, $p<0.001$). The shrunk estimator~\eqref{eq:shrunk} follows the same rationale without committing to a hard decision. It tracks the better of the affine and constant maps within $1\%$ for $N\!\geq\!10$, inherits the significant gains on QMUL-TIMIT in both conditions, and on the corpora where the slope carries no information it degrades gracefully towards the constant predictor rather than towards the affine map, costing $0.07$\,m on VoiceHome2 at $N{=}5$ and matching it thereafter. 


\section{Conclusion}
\label{sec:conclusion}
We studied few-shot calibration of a frozen, synthetic-trained single-channel distance estimator. Zero-shot transfer to real corpora is poor enough that a constant predictor at the corpus mean beats every learned model on all three datasets. Analysing the population risk of the affine map and its one-parameter restrictions shows that the calibrated error depends on how well the estimator correlates with real distances, not its absolute error. Based on the finite-sample cost of each fitted coefficient turns this into a condition, $\rho^2>1/(N+2)$, that accounts for which map wins on which corpus and at which calibration budget, and into a shrinkage estimator that interpolates between the affine and constant maps without a hard threshold. The practical consequence is that the optimal distance estimator should be selected on linear correlation rather than \ac{MAE}, since calibration repairs scale but cannot repair uncorrelated predictions.

\bibliographystyle{IEEEtran}
\bibliography{refs}

\end{document}